\documentclass[floatfix,aps,prl,10pt,twocolumn,superscriptaddress,noeprint]{revtex4-2}
\usepackage{amsmath,bm}
\usepackage{latexsym}
\usepackage{amssymb}
\usepackage{graphics,epstopdf}
\usepackage{hyperref}
\hypersetup{
	colorlinks = true,
	linkcolor =blue,
	citecolor=blue, 
	urlcolor=blue 
}

\newcommand{\splitatcommas}[1]{%
  \begingroup
  \begingroup\lccode`~=`, \lowercase{\endgroup
    \edef~{\mathchar\the\mathcode`, \penalty0 \noexpand\hspace{0pt plus 1em}}%
  }\mathcode`,="8000 #1%
  \endgroup
}
\usepackage[caption=false]{subfig}
\usepackage[percent]{overpic}
\usepackage{comment}
\usepackage{mathtools}
\usepackage{braket}

\newcommand{\mr}[1]{\mathrm{#1}}

\MakeRobust{\eqref}

\newcommand{\Aop}{\hat{A}}
\newcommand{\Bop}{\hat{B}}
\newcommand{\Vop}{\hat{V}}
\newcommand{\phiop}{\hat{\Phi}}
\newcommand{\piop}{\hat{\pi}}
\newcommand{\Oop}{\hat{O}}
\newcommand{\Hop}{\hat{H}}
\newcommand{\Sop}{\hat{S}}
\newcommand{\hrho}{\hat{\rho}}
\newcommand{\Dcal}{\mathcal{D}}

\newcommand{\comm}[2]{\ensuremath{\left[#1,#2\right]}}

\newcommand{\Tr}{\mathrm{Tr}}

\usepackage{soul,xcolor}
\newcommand{\commRS}[1]{{\color{brown}#1}}

\usepackage{microtype}

\begin{document}
\title{Relativistic motion through a thermal bath as a thermodynamic resource}
\author{Rahul Shastri}
\email{rahulkumarkishorbhai.shastri@upol.cz}
\affiliation{Department of Optics, Palacky University, 17. listopadu 1192/12, 779 00 Olomouc, Czech Republic}

\begin{abstract}
We show that a localized quantum system following an arbitrary stationary trajectory and weakly interacting with a stationary thermal bath of a massless scalar field is generically driven into a non-Gibbs steady state by relative motion alone, even without external driving or multiple baths. Relative motion between the system and the bath modifies the standard Kubo–Martin–Schwinger (KMS) relation, preventing relaxation to a Gibbs state. The resulting steady states fall into two distinct classes: (i) nonequilibrium steady states (NESS) with persistent probability currents, and (ii) current-free non-Gibbs steady states characterized by a frequency-dependent effective inverse temperature. We then focus on the simplest stationary trajectory, namely, uniform
relativistic motion with respect to a thermal bath. Using a three-level system as an illustrative example, we demonstrate that the former class can function
as noisy stochastic clocks, while the latter possesses finite nonequilibrium free
energy, enabling work extraction or storage, highlighting their potential as quantum batteries.
\end{abstract}

\maketitle

\emph{Introduction}---Quantum thermodynamics aims to identify resources that enable useful thermodynamic tasks, which require going beyond a single equilibrium thermal bath. It is now well established that nonequilibrium resources such as coherence \cite{Scully2003}, entanglement and correlations \cite{Dillenschneider2009}, squeezed reservoirs \cite{Rossa2014,Klaser2017,Niedenzu2018}, and multiple thermal baths \cite{Uzdin2015} can enable such tasks, underpinning applications in quantum heat engines, quantum batteries, and quantum clocks \cite{Bhattacharjee2021,Campaioli2024,Mitchison2019,Vijit2025}.

Independently, a separate research field—relativistic quantum information (RQI)—has emerged to study the impact of relativistic effects and spacetime structure on quantum systems, motivated by the possibility of harnessing these effects for quantum information processing tasks \commRS{\cite{Mann2012}}. In this context, relativistic motion has been shown to enable entanglement and correlation harvesting \cite{Pozas2015}, magic harvesting \cite{Nyst2025}, etc. Only recently have relativistic effects begun to be explored in thermodynamic settings, including quantum heat engines \cite{Arias2018,Papadatos2021,Barman2022,Mukherjee2022,Ferketic2023,Gallock2023,shaghaghi2025,moustos2025,Moustos2025_2,Hirotani2025,pandit2025}, quantum batteries \cite{mukherjee2024,chen2025}, and quantum metrology \cite{Ahmadi2014,Tian2015,Robles2017,Liu2021,Hossein2023}. From the perspective of quantum thermodynamics, this raises a natural question, whether relativistic effects such as relativistic motion or spacetime structure themselves can be regarded as thermodynamic resources?

In standard open quantum system dynamics, a system weakly coupled to a thermal bath relaxes to a thermal state as a consequence of the KMS condition satisfied by the bath correlation functions. This condition implies detailed balance between excitation and relaxation rates and guarantees thermalization to a Gibbs state. However, in a relativistic setting where the system follows an arbitrary timelike trajectory with respect to the bath, the bath correlation function evaluated along the system trajectory generally fails to satisfy the standard KMS relation \cite{Costa1995,accardi2002,Papadatos2020,Good2020,Perche2021,Bunney2024,Passegger2025}. A particularly important case is that of stationary trajectories, for which the time-translation invariance of a stationary field state is preserved along the
system trajectory. For a quantum field at zero temperature, which is invariant under the full Poincaré group, Letaw identified six inequivalent classes of stationary trajectories, namely inertial motion, uniform linear acceleration, cusp motion, catenary motion, uniform circular motion, and helical motion~\cite{Letaw1981}. In contrast, a finite-temperature field state exhibits a reduced symmetry, being invariant only under time translations, spatial translations, and spatial rotations. This significantly restricts the class of stationary trajectories at finite temperature to inertial motion and uniformly rotating trajectories, such as circular or helical motion~\cite{Costa1995,Bunney2024}. Nonetheless, for any stationary trajectories, the bath correlation function evaluated on the system
trajectory depends only on the proper-time difference, leading to
time-independent transition rates. Moreover, one can introduce a modified KMS relation for bath correlation function, and the resulting excitation and relaxation rates satisfy the detailed balance relation with a frequency-dependent effective inverse temperature \commRS{\cite{Dimitris2019}}. As a result, the steady state reached in the weak-coupling limit is no longer guaranteed to be of Gibbs form. 

In this work, we consider a pointlike quantum system (often referred to as a detector in the RQI literature) interacting with a stationary thermal bath of a massless scalar field while following a stationary trajectory. In the weak-coupling limit, we show that two distinct classes of steady states emerge (i) nonequilibrium steady states with persistent probability currents, and (ii) current-free non-Gibbs steady states characterized by a frequency-dependent effective inverse temperature. Focusing on the simplest stationary trajectory namely uniform relativistic motion with respect to a thermal bath, and using a three-level system as an illustrative example, we demonstrate that the former class can function as noisy stochastic clocks, while the latter possesses finite nonequilibrium free energy, enabling work extraction or storage and highlighting their potential as quantum batteries


\emph{Set-up}---We consider a pointlike quantum system following a timelike trajectory $X(\tau)$, interacting with a massless scalar field with total Hamiltonian
$\Hop=\Hop_{\rm S}+\Hop_{\rm B}+\Vop$,
where $\Hop_{\rm S}=\sum_i\epsilon_i\ket{\epsilon_i}\bra{\epsilon_i}$,
$\Hop_{\rm B}=\frac12\int d^3x\,\big[\piop^2+(\nabla\phiop)^2\big]$,
and the interaction is of Unruh--DeWitt form
$\Vop^{\rm I}(\tau)=\lambda\,\Aop^{\rm I}(\tau)\otimes\phiop[X(\tau)]$,
with the system operator in the interaction picture
$\Aop^{\rm I}(\tau)=\sum_{\omega}e^{-i\omega\tau}\Aop_{\omega}$,
where $\omega$ are the Bohr frequencies of $\Hop_{\rm S}$.
Starting from a factorized initial state and invoking the Born--Markov and secular
approximations, the reduced system dynamics is governed by a
Gorini--Kossakowski--Sudarshan--Lindblad (GKSL) master equation
(see Appendix).
\begin{equation}
\label{eq:GKSLME}
\begin{aligned}
\frac{\partial \hrho}{\partial \tau}
&= -i\comm{\Hop_{\rm S}+\Hop_{\rm LS}}{\hrho} \\
&\quad + \sum_{\omega>0}\!\left[
\Gamma(\omega)\Dcal[\Aop_{\omega}]
+\Gamma(-\omega)\Dcal[\Aop_{\omega}^{\dagger}]
\right]\hrho .
\end{aligned}
\end{equation}
where $\Hop_{\rm LS}=\sum_\omega \Delta(\omega)\Aop_\omega^\dag\Aop_\omega$ is the Lamb-shift Hamiltonian and
$\Dcal[\Oop]\hrho=\Oop\hrho\Oop^\dag-\frac12\{\Oop^\dag\Oop,\hrho\}$,
with $\{\Aop,\Bop\}=\Aop\Bop+\Bop\Aop$.
The dissipative dynamics in Eq.~\eqref{eq:GKSLME} is controlled by the bath
spectral function
\begin{equation}
\Gamma(\omega)=\lambda^2\!\int_{-\infty}^{\infty}\!d\tau\,e^{i\omega\tau}G(\tau,0),
\label{eq:bathSpectralFunc}
\end{equation}
which is time independent for stationary trajectories. For such trajectories with nonvanishing transition rates, one may introduce a modified KMS 
relation~\cite{Dimitris2019},
\begin{equation}
\frac{\Gamma(\omega;\alpha_i)}{\Gamma(-\omega;\alpha_i)}
= e^{\beta_{\rm eff}(\omega;\alpha_i)\,\omega},
\label{eq:GDBC}
\end{equation}
where $\beta_{\rm eff}(\omega;\alpha_i)$ is a frequency-dependent effective inverse
temperature and $\alpha_i$ denote parameters characterizing the trajectory (e.g.,
velocity, acceleration, or radius).
\begin{figure}
\subfloat{\begin{overpic}[width=0.8\linewidth]{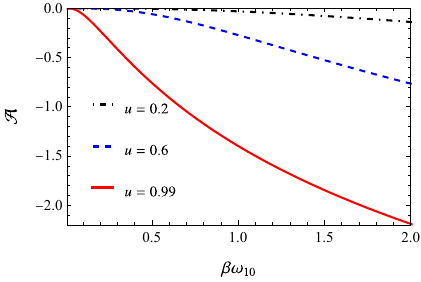}
	\put(80,60){\textbf{(a)}}
	\end{overpic}
	} \\
\subfloat{\begin{overpic}[width=0.8\linewidth]{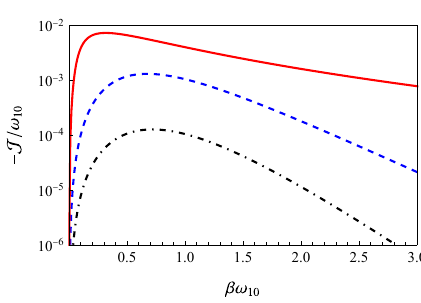}
	\put(80,60){\textbf{(b)}}
	\end{overpic}
	}
 \caption{(Color Online) Affinity $\mathcal{A}$ (a) and current $-\mathcal{J}$ (b) of three level system, as function of inverse temperature $\beta$ for three different value of velocity small $u=0.2$ (black dot-dashed), intermediate $u=0.6$ (blue dashed) and ultrarelativistic $u=0.99$ (red solid). Other parameter values are $\omega_{10}=1.0, \omega_{21}=3.1$. }
\label{fig:fig1}
\end{figure}

\emph{Results}-- In the long-time limit of Eq.~\eqref{eq:GKSLME}, coherences in the energy eigenbasis vanish, and the steady state becomes diagonal (see Appendix). The populations $p_i=\bra{\epsilon_i}\hrho\ket{\epsilon_i}$ then obey a Pauli master equation,
\begin{align}
\dot{p_i}  = -\sum_j   J_{i\to j},
\end{align}
where $ J_{i\to j} = p_i k_{i\to j} - p_j k_{j\to i}$ denotes the probability current  with the transition rates,
\begin{align}k_{i\to j} = |\bra{\epsilon_i}\Aop\ket{\epsilon_j}|^2\Gamma(\omega_{ij};\alpha_i),
\label{eq:ratesMatrix}
\end{align}
with $\omega_{ij} = \epsilon_i-\epsilon_j$. The steady-state populations satisfy $\dot p_i = \sum_j J_{i\to j} = 0$. However, this condition does not in general imply that the individual currents $J_{i\to j}$ vanish. We find that the steady states fall into two distinct classes. If, for every closed sequence of allowed transitions
$\epsilon_{i_1}\!\to\!\epsilon_{i_2}\!\to\!\cdots\!\to\!\epsilon_{i_n}\!\to\!\epsilon_{i_1}$
with nonvanishing rates, Kolmogorov’s loop condition~\cite{Schnakenberg1976}
\begin{align}
\sum_{r=1}^{n}\ln\frac{k_{i_r\to i_{r+1}}}{k_{i_{r+1}\to i_r}}=0,
\qquad i_{n+1}\equiv i_1,
\label{eq:DBC2}
\end{align}
is satisfied, then the ratio of transition rates can be written as
$\ln(k_{i\to j}/k_{j\to i}) = F_i - F_j$ and the steady-state populations satisfy detailed balance $\frac{p_j}{p_i}
= e^{\beta_{\rm eff}(\omega_{ij};\alpha_i)\,\omega_{ij}}$,
implying that all probability currents vanish. Consequently, the steady-state assumes the exponential form
\begin{align}
\hat{\rho}^{\mathrm{ss}}
= \frac{1}{Z_F}\sum_i e^{-F_i}\ket{\epsilon_i}\bra{\epsilon_i},
\qquad
Z_F := \sum_i e^{-F_i},
\label{eq:ExpForm}
\end{align}
where $F_i$ are obtained by fixing an arbitrary reference level $i_0$ and setting $F_{i_0}=0$. If the loop condition~\eqref{eq:DBC2} is violated for at least one allowed closed
cycle of transitions, the system relaxes to a NESS with persistent probability currents. In this case, no closed-form
expression of the form Eq.~\eqref{eq:ExpForm} exists, and the steady state must instead
be obtained by solving $\sum_j J_{i\to j}=0$. Note that while checking the loop condition Eq.~\eqref{eq:DBC2}, situations may arise in which no
closed sequence of allowed transitions exists due to vanishing matrix elements
of the system operator entering the rates Eq.~\eqref{eq:ratesMatrix}, e.g., because
of symmetries or the specific system--bath coupling. In such cases, probability
currents are necessarily absent, and the steady state still admits the
exponential form Eq~\eqref{eq:ExpForm}.

\begin{figure}
\subfloat{\begin{overpic}[width=0.8\linewidth]{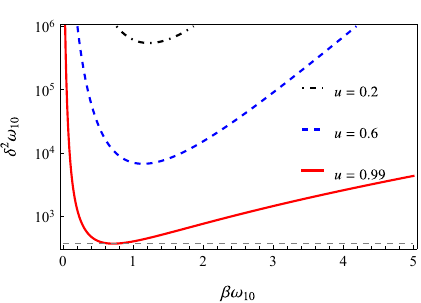}
 \put(85,55){\textbf{(a)}}
 \end{overpic}
 } \\
\subfloat{\begin{overpic}[width=0.8\linewidth]{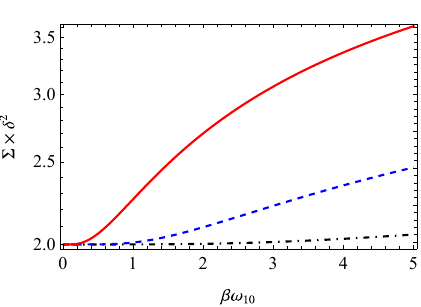}
 \put(85,55){\textbf{(b)}}
 \end{overpic}
 }
 \caption{(Color Online) Relative uncertainty $\delta^2$ (a) and product of relative uncertainty with entropy production $\mathcal{\delta}^2\Sigma$ (b) as a function of inverse temperature $\beta$ for different values of velocity $u$. Other parameter values are same as Fig.~\eqref{fig:fig1}.}
\label{fig:fig2}
\end{figure}

Interestingly, the steady state~\eqref{eq:ExpForm} is generally non-Gibbsian due
to the frequency dependence of the effective inverse temperatures
$\beta^{\rm eff}(\omega;\alpha_i)$. For multilevel systems with distinct
transition frequencies, this can prevent thermal populations and
may render the steady state thermodynamically non-passive. In particular, population inversion may
occur for transitions with $\beta^{\rm eff}(\omega;\alpha_i)<0$.  To best of our knowledge, no general argument ensures $\beta^{\rm eff}(\omega;\alpha_i)\ge0$
for arbitrary stationary trajectories in Eq.~\eqref{eq:GDBC}, although it is
non-negative in the few stationary cases studied~\cite{Letaw1981,Good2020}.
More subtly, even when $\beta^{\rm eff}(\omega;\alpha_i)>0$, population inversion
may still arise if the product $\beta^{\rm eff}(\omega;\alpha_i)\,\omega$ has non-trivial frequency dependence which can lead to $F_j<F_i$ for $\epsilon_j>\epsilon_i$ for some $i,j$ in Eq.~\eqref{eq:ExpForm}. A two-level system which has only a single Bohr frequency $\omega$ is a trivial case for which the steady state can always be written in the Gibbs form with a unique $\beta^{\rm eff}(\omega)$.

As a concrete example, we consider the field in a stationary thermal
state formally written as $\hrho_{\rm B}=e^{-\beta\Hop_{\rm B}}/\Tr_{\rm B}(e^{-\beta\Hop_{\rm B}})$ at
inverse temperature $\beta=1/T$, and a stationary trajectory corresponding
to inertial motion with uniform velocity $u$,
$X(\tau) = \left(\frac{1}{\sqrt{1-u^2}}, \frac{u}{\sqrt{1-u^2}},0,0\right)\tau$ (units $c=1$). In this case, the bath spectral function Eq.~\eqref{eq:bathSpectralFunc} admits a
simple analytical expression (see Appendix). First, to demonstrate the NESS with persistent probability currents, we consider a three-level system with an interaction operator
\begin{equation}
\Aop^{\rm I}(\tau)
=\sum_{(ij)\in\{10,21,20\}}\lambda_{ij}
\bigl(\Sop_{ij}e^{-i\omega_{ij}\tau}
+\Sop_{ij}^\dag e^{i\omega_{ij}\tau}\bigr),
\label{eq:Threelevelinteraction}
\end{equation}
where we define the lowering and raising operators,
\begin{align}
\Sop_{10}=\ket{\epsilon_1}\bra{\epsilon_0},\quad
\Sop_{21}=\ket{\epsilon_2}\bra{\epsilon_1},\quad
\Sop_{20}=\ket{\epsilon_2}\bra{\epsilon_0},
\end{align}
satisfying $[H_S,S_{ij}]=-\omega_{ij}S_{ij}$ with Bohr frequencies
$\omega_{ij}=\epsilon_i-\epsilon_j>0$ and $S_{ji} = S_{ij}^\dag$. Note that we assume non-degenerate and well-separated Bohr frequencies such that the secular approximation is valid. The NESS with persistent current can be characterized by the cycle affinity,
\begin{equation}
\begin{alignedat}{2}
\mathcal{A}
&=\ln\frac{k_{2\to0}k_{0\to1}k_{1\to2}}{k_{0\to2}k_{1\to0}k_{2\to1}}\\
&=\beta_{\rm eff}(\omega_{20})\omega_{20}
-\beta_{\rm eff}(\omega_{21})\omega_{21}
-\beta_{\rm eff}(\omega_{10})\omega_{10}.
\end{alignedat}
\end{equation}
\noindent For $u=0$, $\beta_{\rm eff}(\omega;\beta,u)\equiv\beta$, implying $\mathcal A=0$. For $u\neq 0$, one generically finds $\mathcal A\neq 0$ and a nonzero steady-state probability current. For a three-level system, the steady-state currents along each transition are equal,
\begin{equation}
\mathcal J \equiv J_{0\to1}^{\rm ss} = J_{1\to2}^{\rm ss} = J_{2\to0}^{\rm ss}.
\end{equation} In addition, to maintain the system in NESS, we also have a non-zero entropy production rate $\Sigma=\mathcal J\,\mathcal{A}\ge 0$. We adopt the convention that $\mathcal J>0$ denotes current along $0\to1\to2\to0$. Also the $\operatorname{sign}(\mathcal J)=\operatorname{sign}(\mathcal{A})$. 
In the small-velocity limit $u\ll 1$,
one finds,
\begin{align}
    \mathcal{A} &\approx \frac{\beta^2u^2}{6}\left[\omega_{10}^2\coth\left(\frac{\omega_{10}\beta}{2}\right)+\omega_{21}^2\coth\left(\frac{\omega_{21}\beta}{2}\right) \right. \nonumber\\
     &- \left. (\omega_{10}+\omega_{21})^2\coth\left(\frac{(\omega_{10}+\omega_{21})\beta}{2}\right) 
    \right],
\end{align}
where $\omega_{20}=\omega_{21}+\omega_{10}$. Moreover, using the inequality,
\begin{equation}
(a+b)^2 \coth\!\left(\frac{(a+b)\beta}{2}\right)
\ge a^2 \coth\!\left(\frac{a\beta}{2}\right)
+ b^2 \coth\!\left(\frac{b\beta}{2}\right).
\end{equation}
\noindent for $a,b\geq0$, we see that $\mathcal{A}\le 0$ to order $u^2$. Therefore, for small $u$ the current circulates opposite to the chosen convention with $\mathcal J<0$ (i.e., the current flows along $0\to2\to1\to0$ ). 

In Fig.~\ref{fig:fig1}(a) we see that, the affinity $\mathcal{A}$ vanishes as $u\to0$ and becomes increasingly negative with increasing $u$, reaching its largest magnitude in the low-temperature and ultrarelativistic regimes. The sign reflects the direction of the steady-state current and agrees with the small-velocity analysis, and persists even beyond that. Fig.~\ref{fig:fig1}(b) shows the corresponding current $\mathcal{J}$, whose magnitude grows from zero at high temperatures, peaks at an intermediate inverse temperature, and decays again as $\beta\to\infty$. In the weak-driving regime $\mathcal{A}\ll1$, the current is approximately linear, $\mathcal{J}\simeq K_0\mathcal{A}$, with $K_0$ depending only on $\beta$ and the Bohr frequencies. Since $\mathcal{A}\propto u^2$ for $u\ll1$, this implies $\mathcal{J}\propto u^2$ at low velocities, with an optimal $\beta$ set by the system’s Bohr frequencies.

\emph{Stochastic clock}-- We show that the three-level NESS with persistent current can function as a stochastic clock \cite{Barato2016,Erker2017}. A clock cycle is defined by the net transition $2\to0$, with one tick corresponding to one complete cycle (see Supplemental Material). The stochastic variable $n(\tau)$ counts these transitions, increasing by $+1$ for $2\to0$ and decreasing by $-1$ for $0\to2$. In the long-time limit, $\langle n\rangle_\tau \simeq \mathcal{J}\tau$ and $\mathrm{Var}(n)_\tau \simeq 2D\tau$, where $D$ the diffusion constant. The clock performance is characterized by the ticking rate and relative uncertainty,
\begin{align}
\mathcal{J} = \frac{\langle n \rangle_\tau}{\tau}, \quad \quad \delta^2 = \frac{2D}{\mathcal{J}^2},
\end{align}
where the steady-state current $\mathcal{J}$ is interpreted as the number of cycles per unit time and $\delta^2$ characterizes relative uncertainty per unit time. In time interval $\tau$, we expect to read $\mathcal{J}\tau$ number of cycles on average with relative uncertainly $\delta\frac{1}{\sqrt{\tau}}$. 

Fig.~\ref{fig:fig2}(a) shows that the relative uncertainty $\delta^2$ exhibits a minimum at an intermediate inverse temperature $\beta$, with $\delta^2 \sim 10^2\omega_{10}^{-1}$, indicating substantial fluctuations. The entropy production rate $\Sigma$ is constrained by the thermodynamic uncertainty relation (TUR) $\delta^2\Sigma\ge2$, and Fig.~\ref{fig:fig2}(b) shows that the product $\delta^2\Sigma$ remains well above this bound, approaching it only close to equilibrium ($u\to0$, $\mathcal{A}\to0$). This reflects the dissipation–precision tradeoff and indicates that entropy production is not efficiently converted into improved clock precision. Overall, the system functions as a stochastic clock with a finite ticking rate but significant noise, although the error decreases as $1/\sqrt{\tau}$ at long times.
\begin{figure}[h!]
    \centering
\includegraphics[width=0.8\linewidth]{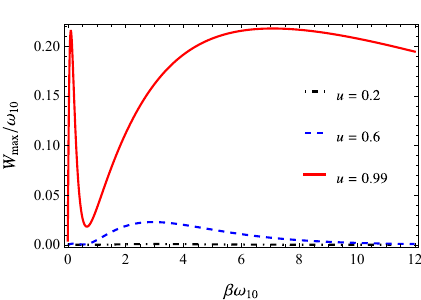}
    \caption{(Color Online) Plot of maximum extractable work $\mathcal{W}_\mr{max}$ of steady-state $\hrho^\mr{ss}$ for different values of velocity $u$. Other parameter values are same as Fig.~\eqref{fig:fig1}}
 \label{fig:fig4}
\end{figure}

\emph{Quantum battery}-- Consider a general multilevel system undergoing uniform relativistic motion and satisfying the loop condition in Eq.~\eqref{eq:DBC2}. We find that the resulting steady state possesses finite nonequilibrium free energy despite having vanishing ergotropy. This follows from two observations. First, the steady state is diagonal in the energy eigenbasis, so any nonzero ergotropy would require population inversion. Second, such inversion is excluded because the effective inverse temperature satisfies $\beta_{\rm eff}(\omega,\beta,u) \ge 0$, and the product $\beta_{\rm eff}(\omega)\,\omega$ increases monotonically with $\omega$ (see Appendix), ensuring monotonically decreasing populations with energy. Although no work can be extracted via unitary operations, the non-Gibbs nature of the steady state allows work extraction in the form of nonequilibrium free energy, either through interaction with an external bath or by using multiple copies of the system~\cite{Manzano2018,Alicki2013}. 

Let us illustrate this explicitly for a three-level system. Setting $\lambda_{01}=0$ in Eq.~\eqref{eq:Threelevelinteraction} allows only the $2\leftrightarrow 0$ and $2\leftrightarrow 1$ transitions while $1\leftrightarrow 0$ is forbidden. In this case in steady-state each individually current vanishes i.e. $J_{2\to0}^{\rm ss}=J_{1\to2}^{\rm ss}=0$, and the steady state $\hrho^\mr{ss}$ is of the form Eq.~\eqref{eq:ExpForm} with $F_0=0$, $F_1=\beta_{\rm eff}(\omega_{20})\omega_{20}
-\beta_{\rm eff}(\omega_{21})\omega_{21}$,
$F_2=\beta_{\rm eff}(\omega_{20})\omega_{20}$,
and $Z_F=\sum_{i=0}^{2}e^{-F_i}$.
The maximum extractable work is quantified by the non-equilibrium free energy difference between the given steady-state and the reference Gibbs thermal state. 
\begin{align}
    \mathcal{W}_\mr{max} = \mathcal{F}(\hrho^\mr{ss}) - \mathcal{F}(\hrho^\beta) = \frac{1}{\beta}D\left(\hrho^\mr{ss}||\hrho^\beta\right),
    \label{eq:wmax}
\end{align}
where $\mathcal{F}(\hrho) = \mr{Tr}(\Hop_\mr{S}\hrho) - \frac{1}{\beta}S(\hrho)$, $S(\hrho) = -\mr{Tr}\left[\hrho\ln \hrho\right]$ is von Neumann entropy, $\hrho^\beta = e^{-\beta \Hop_\mr{S}}/{\mr{Tr}e^{-\beta \Hop_\mr{S}}}$ is reference Gibbs thermal state at bath temperature in its rest frame $\beta$, and $D\left(\hrho||\hat{\sigma}\right) = \mr{Tr}\left[\hrho(\ln \hat{\sigma}-\ln \hrho)\right]$ is quantum relative entropy.
We define the reference temperature as the bath rest frame temperature $\beta$. Although one might attempt to define a temperature in the comoving frame of the system, no unique inverse temperature can be assigned to a relativistically moving bath \cite{Papadatos2020}. 

Fig.~\ref{fig:fig4} shows that the extractable work vanishes in both the high ($\beta\to0$) and low ($\beta\to\infty$) temperature limits, with a pronounced peak followed by a broad maximum at intermediate temperatures. This behavior originates from the deviation between the effective inverse temperature $\beta_{\mathrm{eff}}(\omega;\beta,u)$ associated with the largest energy gap and the bath inverse temperature $\beta$. At high temperatures the steady state approaches a maximally mixed state, while at low temperatures it approaches the ground state, suppressing work extraction in both limits. At intermediate temperatures, crossings between $\beta_{\mathrm{eff}}$ and $\beta$ alternately enhance and suppress work extraction, giving rise to the observed features.

\emph{Discussion}--To illustrate  NESS with nonzero affinity, we considered a three-level system in $\Delta$ configuration. While such configurations are not generic for natural atoms due to symmetry and dipole selection rules, they can arise in chiral molecules~\cite{Ye2018} and can be engineered in artificial atoms such as superconducting circuits~\cite{Liu2005}.

Turning to the energetic origin of this resource, maintaining a NESS with probability currents requires a continuous consumption of
nonequilibrium resources, which in the present setting originate from the
energy cost required to sustain the relative motion between the system and the
bath. This continuous energy input is dissipated into the bath and can be connected to
quantum friction effects associated with the relative motion between a quantum
system and its environment, such as Einstein--Hopf drag~\cite{Lach2012}. By contrast, in the current-free case $\mathcal{A}=0$, no entropy production is
required to maintain the steady state, and any extractable work must have been
injected during the transient dynamics that drive the system away from Gibbs
equilibrium.

\emph{Conclusion}--We have shown that stationary motion of a quantum system relative to a single thermal bath generically leads to non-Gibbs steady states, arising from a modification of the KMS condition. For uniform relativistic motion, illustrated using a three-level system, these steady states can either sustain NESS with persistent probability currents, enabling noisy stochastic clocks, or store finite nonequilibrium free energy, functioning as quantum batteries. Our results identify relativistic motion as a genuine thermodynamic resource and motivate the exploration of nonequilibrium thermodynamic tasks induced by motion and spacetime structure.

\medskip
\noindent\emph{Acknowledgements.}--The author thanks B. P. Venkatesh, K. Adhikary, and V. Singh for useful comments.

\vspace{0.1in}

\bigskip
\noindent \emph{Appendix A: Details on GKSL master equation}--- For the setup described in the main text, Starting from a factorized initial state $\hrho_{\mathrm{tot}}(0)=\hrho(0)\otimes\hrho_{\mathrm{B}}$ and within the weak-coupling Born–Markov approximation, tracing out the field degrees of freedom yields the interaction-picture master equation to second order in the coupling,
\begin{align}
    \frac{\partial \hrho^\mr{I}(\tau)}{\partial \tau} &= -\lambda^2 \int_0^{\infty} ds \comm{\Aop^\mr{I}(\tau)}{\Aop^\mr{I}(s)\hrho^\mr{I}(\tau)} G(\tau,s) + \mr{h.c},
\end{align}
where,
\begin{align}
 G\left(\tau,s\right) = \Tr_\mr{B}\left(\phiop(X(\tau))\phiop(X(s))\hrho_\mr{B}\right),
\end{align}
is the field two-point (Wightman) correlation function in the state $\rho_{\mathrm B}$, evaluated along the system trajectory. We focus on stationary trajectories, for which the Wightman function depends only on the proper-time difference, $G(\tau,s)=G(\tau-s,0)$. Using the decomposition of the system operator
$\Aop^{\mathrm I}(\tau)
= \sum_{\omega} e^{-i\omega \tau}\, \Aop_{\omega}$, the equation of motion for the system in the interaction picture reads
\begin{align}
\frac{\partial \hrho^{\mathrm I}(\tau)}{\partial \tau}
&= \lambda^{2}
\sum_{\omega,\omega'}
\tilde g(\omega')\,
e^{-i(\omega-\omega')\tau}
\nonumber\\
&\quad\times
\Big(
\Aop_{\omega}\,\hrho^{\mathrm I}(\tau)\,\Aop_{\omega'}^{\dagger}
-
\Aop_{\omega'}^{\dagger}\Aop_{\omega}\,\hrho^{\mathrm I}(\tau)
\Big)
+ \mathrm{h.c.},
\end{align}
where the coefficient $\tilde g(\omega)$ is given by the one-sided Fourier transform of the field correlation function,
\begin{align}
\tilde g(\omega)
= \int_{0}^{\infty} \! ds \, e^{i\omega s}\, G(s,0).
\end{align}
The operators $\Aop_{\omega}$ are defined as
$\Aop_{\omega}
= \sum_{p,k:\,\epsilon_{p}-\epsilon_{k}=\omega}
\bra{\epsilon_{k}}\Aop\ket{\epsilon_{p}}
\ket{\epsilon_{k}}\bra{\epsilon_{p}},
$
where $\Hop_{\mathrm S}\ket{\epsilon_{k}}=\epsilon_{k}\ket{\epsilon_{k}}$,
with the convention that positive Bohr frequencies $\omega$ correspond to downward (emission) transitions of the system, while negative $\omega$ correspond to upward (absorption) transitions.
These operators satisfy $[\Hop_{\mathrm S}, \Aop_{\omega}] = -\omega\,\Aop_{\omega}$,
$[\Hop_{\mathrm S}, \Aop_{\omega}^{\dagger}] = \omega\,\Aop_{\omega}^{\dagger}$,
and consequently
$[\Hop_{\mathrm S},\Aop_{\omega}\Aop_{\omega}^{\dagger}]
=
[\Hop_{\mathrm S},\Aop_{\omega}^{\dagger}\Aop_{\omega}]
= 0$.
To simplify the equation, we invoke the secular approximation, neglecting rapidly oscillating terms with $\omega\neq\omega'$. This is valid when the Bohr frequencies are well separated and $|\omega-\omega'|^{-1}$ is short compared to the system’s relaxation timescale. Transforming back to the Schrödinger picture then yields,
\begin{align}
    \frac{\partial \hrho(\tau)}{\partial \tau} &= -i\comm{\Hop_\mr{S}}{\hrho(\tau)} \\ \nonumber 
    &+ \lambda^2 \sum_{\omega}\Tilde{g}(\omega) \left(\Aop_\omega \hrho(\tau)\Aop_{\omega}^\dag - \Aop_{\omega}^\dag\Aop_\omega \hrho(\tau)\right) + \mr{h.c}.
\end{align}
We split $\Tilde{g}(\omega)$ into its real and imaginary part as,
\begin{align}
    \lambda^2\Tilde{g}(\omega) = \frac{1}{2}\Gamma(\omega) + i\Delta(\omega),
    \label{eq:WightmanTransform}
\end{align}
with which we get Eq.~\eqref{eq:GKSLME} of the main text. Further details on the derivation and validity of the master equation can be found in
Refs.~\cite{Dimitris2019,Papadatos2020,Han2025}. Note that the form of Eq.~\eqref{eq:GKSLME} is that of the standard GKLS master equation for thermalisation but with bath spectral function satisfying generalized detailed balance relation Eq.~\eqref{eq:GDBC}. In the energy eigenbasis $\Hop_\mr{S} \ket{\epsilon_i} = \epsilon_i \ket{\epsilon_i}$, the off-diagonal
elements $\rho_{ij} = \bra{\epsilon_i}\hrho\ket{\epsilon_j}$ for $i\neq j$ with $\omega_{ij} = \epsilon_i-\epsilon_j$ obey,
\begin{align}
    \dot\rho_{ij}
    &= -i\big[\omega_{ij} + \Delta(\omega_{ij})\big]\rho_{ij}- \frac{1}{2}\sum_l \big( k_{l\to i} + k_{l\to j}\big)\rho_{ij},
\end{align}
where $\Delta(\omega_{ij})$ comes from the Lamb-shift Hamiltonian. For $\sum_l \bigl(k_{l\to i}+k_{l\to j}\bigr) > 0 \,\text{for all } i\neq j$,
all coherences decay to zero in the long-time limit, and the steady state is therefore diagonal in the energy eigenbasis.

For the particular example of the field to be in a stationary thermal Gibbs state which is completely characterized by its Wightman function,
\begin{align}
G^{\beta}\!\left(X(\tau),X(\tau')\right)
= - \frac{1}{4\pi^{2}}
\sum_{n=-\infty}^{+\infty}
\Big[
\left(t-t'-in\beta-i\varepsilon\right)^{2}
\nonumber\\
\qquad\qquad
- \left|\mathbf{x}-\mathbf{x}'\right|^{2}
\Big]^{-1},
\label{eq:thermalWightman}
\end{align}
where $X(\tau)=(t,\mathbf{x})$ and $X(\tau')=(t',\mathbf{x}')$ denote the system trajectory. We focus on a particular choice of \emph{stationary trajectory}  that is system undergoing inertial motion with uniform velocity $u$ along the trajectory
$X(\tau)=\left(\frac{1}{\sqrt{1-u^{2}}},\frac{u}{\sqrt{1-u^{2}}},0,0\right)\tau$,
where $0\le u<1$ is the velocity in units where $c=1$.
In this case, the bath spectral function entering the master equation reads \cite{Papadatos2020}
\[
\Gamma(\omega;\beta,u)=
\begin{cases}
\gamma(\omega)\!\left[N(\omega,\beta,u)+1\right], & \omega>0,\\
\gamma(|\omega|)N(|\omega|,\beta,u), & \omega<0,
\end{cases}
\]
with $\gamma(\omega)=\lambda^{2}\omega/(2\pi)$ and
\begin{align}
N(\omega,\beta,u)
= \frac{\sqrt{1-u^{2}}}{2\beta \omega u}
\ln\!\left(
\frac{1-e^{-\beta\omega \sqrt{\frac{1+u}{1-u}}}}
{1-e^{-\beta\omega \sqrt{\frac{1-u}{1+u}}}}
\right).
\end{align}
For this the frequency dependent effective inverse temperature takes the form
\begin{align}
\beta_{\mathrm{eff}}(\omega,\beta,u)
= -\frac{1}{\omega}
\ln\!\left(\frac{N(\omega,\beta,u)}{N(\omega,\beta,u)+1}\right).
\end{align}
Note that since $N(\omega,\beta,u)\geq 0$, which implies
\begin{align}
0<\frac{N}{N+1}<1
\quad\Rightarrow\quad
\beta_{\rm eff}(\omega,\beta,u)
=-\frac{1}{\omega}\ln\!\left(\frac{N}{N+1}\right)\ge 0.
\end{align}
Moreover,since
\[
\omega\,\beta_{\rm eff}(\omega,\beta,u)=\ln\!\left(1+\frac{1}{N(\omega,\beta,u)}\right),
\]
and $N(\omega,\beta,u)$ is strictly increasing with $\omega$, it follows that $\omega\,\beta_{\rm eff}(\omega,\beta,u)$ is strictly increasing function of $\omega$.

\bibliography{mybib}
\end{document}